\documentclass[twocolumn,twocolappendix]{aastex631}

\usepackage{amsmath}
\usepackage{hyperref}

\begin{document}

\title{Richardson--Lucy deconvolution with a spatially Variant point-spread function of Chandra: Supernova Remnant Cassiopeia A as an Example}

\author[0000-0002-5809-3516]{Yusuke Sakai}
\affiliation{Department of Physics, Rikkyo University, Toshima-Ku, Tokyo, 171-8501, Japan}

\author[0000-0003-4808-893X]{Shinya Yamada}
\affiliation{Department of Physics, Rikkyo University, Toshima-Ku, Tokyo, 171-8501, Japan}

\author[0000-0001-9267-1693]{Toshiki Sato}
\affiliation{Department of Physics, Rikkyo University, Toshima-Ku, Tokyo, 171-8501, Japan}
\affiliation{Department of Physics, School of Science and Technology, Meiji University, 1-1-1 Higashi Mita, Tama-ku, Kawasaki, Kanagawa 214-8571, Japan}

\author[0000-0002-3752-0048]{Ryota Hayakawa}
\affiliation{Department of Physics, Rikkyo University, Toshima-Ku, Tokyo, 171-8501, Japan}
\affiliation{International Center for Quantum-field Measurement Systems for Studies of the Universe and Particles (QUP), KEK, 1-1 Oho, Tsukuba, Ibaraki 305-0801, Japan}
\author[0000-0001-6409-7735]{Ryota Higurashi}
\affiliation{Department of Physics, Rikkyo University, Toshima-Ku, Tokyo, 171-8501, Japan}

\author[0000-0001-8335-1057]{Nao Kominato}
\affiliation{Department of Physics, Rikkyo University, Toshima-Ku, Tokyo, 171-8501, Japan}

\begin{abstract}
Richardson--Lucy (RL) deconvolution is one of the classical methods widely used in X-ray astronomy and other areas. Amid recent progress in image processing, RL deconvolution still leaves much room for improvement under a realistic situations. One direction is to include the positional dependence of a point-spread function (PSF), so-called RL deconvolution with a spatially variant PSF (RL$_{\rm{sv}}$). Another is the method of estimating a reliable number of iterations and their associated uncertainties. We developed a practical method that incorporates the RL$_{\rm{sv}}$~algorithm and the estimation of uncertainties. As a typical example of bright and high-resolution images, the Chandra X-ray image of the supernova remnant Cassiopeia~A was used in this paper. RL$_{\rm{sv}}$ deconvolution enables us to uncover the smeared features in the forward/backward shocks and jet-like structures. We constructed a method to predict the appropriate number of iterations by using statistical fluctuation of the observed images. Furthermore, the uncertainties were estimated by error propagation from the last iteration, which was phenomenologically tested with the observed data. Thus, our method is a practically efficient framework to evaluate the time evolution of the remnants and their fine structures embedded in high-resolution X-ray images.
\end{abstract}

\section{Introduction}
Imaging analysis is critically important for studying diffuse celestial sources.
X-ray astronomy, starting with the first space application of X-ray CCD in ASCA \citep{burke1994ccd}, 
has delivered detailed images of various celestial objects; 
e.g., supernova remnants such as SN1006 \citep{bamba2003small} and Cassiopeia A \citep[hereafter Cas~A;][]{hwang2004million}, 
and galaxy clusters such as A2142 \citep{markevitch2000chandra}.
Since the X-ray mirrors used in Chandra are the largest and most precisely built, 
exceeding the angular resolution of Chandra is considered to be challenging. 
Therefore, enhancing the technique of imaging analysis has been an essential direction 
to utilize the highest spatial resolution and data accumulated over decades. 

X-rays are collected primarily by total reflection from the surface of an X-ray mirror, therefore the response function for the distribution of focused X-rays, called the point-spread function (PSF), is nearly energy independent.
On the condition that a PSF is independent of incoming photon energy and the position of the focal plane, 
a reverse calculation of a convolution of PSF, 
so-called image deconvolution (see the review on deconvolution in astronomy by \cite{starck2002deconvolution}), is highly simplified.
There are various deconvolution methods proposed by assuming that a PSF is constant during the deconvolution process, 
such as the deconvolution of Suzaku XIS \citep{sugizaki2009deconvolution}. 
One of the latest examples is the image restoration algorithm Expectation via Markov chain Monte Carlo \citep{esch2004image}. 
It is applied to the double active galactic nuclei in NGC 6240 \citep[e.g.,][]{fabbiano2020revisiting,paggi2022dissecting}, 
succeeding in finely resolving the two cores. Similarly, a classical method, Richardson--Lucy (RL) deconvolution proposed by \cite{richardson1972bayesian} and \cite{lucy1974iterative}, is  often used \citep[e.g.,][]{grefenstette2015locating,thimmappa2020chandra,sobolenko2022ngc}. 

The choice of method depends on the trade-off between accuracy and computational cost. 
Relaxing the condition that a PSF is positional-independent and/or energy-independent, the deconvolution methods increase the complexity of the calculation. 
RL deconvolution is one of the simplified methods but still has room for improvement in practical situations. 
In gamma-ray astronomy, a PSF can change by one order of magnitude with energy and incident angle; 
it is calculated for each event, e.g., RL algorithm optimized for Fermi-LAT and EGRET \citep{tajima2007studies}. 
In contrast, as the number of photons is much larger in X-ray astronomy, event-by-event reconstruction is less practical; image-based reconstruction thus can be the first choice in X-rays. 
However, there are few studies on extending the RL method, especially their application to diffuse sources obtained by Chandra. 
We therefore explored its applicability to the Chandra data and considered the associated systematic errors. 

In this paper, we implement RL deconvolution with a spatially variant PSF (RL$_{\rm{sv}}$) algorithm, assuming it to be used for Chandra images. 
Section~\ref{Method} describes the principle of the RL$_{\rm{sv}}$~method. 
One of the technical difficulties is reducing computational cost in calculating PSFs. 
This is solved by decimating the sampling interval of PSFs, while a side-effect is discussed in Section~\ref{Enhancement Technique and Possibilities}. 
Section~\ref{Application to Observed Data} presents an example of its application to a diffuse source observed by Chandra. 
We apply the RL$_{\rm{sv}}$ method to the supernova remnant of Cas~A as an example,  
because Cas~A is bright and extended over the entire field of view of the ACIS detector, which would be the best target for the first application.
The remnant is intensively studied because of its unique structure and evolution, e.g., the velocities and thickness of shocked filaments \citep{patnaude2009proper, sato2018x, tsuchioka2022x}, 
where the method can contribute to advancing our understanding of the phenomena. 
In Section~\ref{Uncertainty Estimation}, we propose a reliable number of stop iterations and uncertainties of the method. 
We develop the method to estimate the number of convergent iterations by generating fluctuations due to statistical errors during iteration. 
Furthermore, the uncertainty on the RL$_{\rm{sv}}$-deconvolved image is estimated by using the law of error propagation \citep[e.g.,][]{ku1966notes}. 
As a result, filaments and ambiguous structures of Cas~A are deconvolved to be sharper with some knowledge of the statistical uncertainties. 

\begin{table*}[ht!]
 \caption{Basic information on the Chandra Observations of Cas~A Used in this Paper}
  \centering
   \begin{tabular}{ccccccc} \hline\hline
      Obs. ID & Obs. Start & Exp. Time &  detector & R.A. & Decl. & Roll\\
       & yyyy~mmm~dd & (ks) & & (deg) & (deg) & (deg)\\ \hline
     4636 & 2004 Apr 20 & 143.48 & ACIS-S & 350.9129 & 58.8412 & 49.7698\\
     4637 & 2004 Apr 22 & 163.50 & ACIS-S & 350.9131 & 58.8414 & 49.7665\\
     4639 & 2004 Apr 25 & 79.05 & ACIS-S & 350.9132 & 58.8415 & 49.7666 \\
     5319 & 2004 Apr 18 & 42.26 & ACIS-S & 350.9127 & 58.8411 &  49.7698\\ \hline
     5196 & 2004 Feb 8 & 49.53 & ACIS-S & 350.9129 & 58.7933 &  325.5035 \\ \hline
   \end{tabular}
  \label{casA_data}
\end{table*}

\section{Method}\label{Method}
\subsection{RL Deconvolution}
The RL~algorithm iteratively estimates a true image from an observed image using Bayesian inference.
It generally assumes that the PSF does not change with a position in the image. 
The RL~algorithm is expressed by 
\begin{equation}
W_{i}^{(r+1)} = W_{i}^{(r)} \sum_k \frac{P_{ik}H_k}{\sum_j P_{jk}W_{j}^{(r)}}, 
\label{rl_eq}
\end{equation}
where $i$ and $j$ are mapping the image in the sky, and $k$ is mapping the image on the detector.
The indices of the summation run through all the pixels. 
$W^{(r)}$ is the restored image after $r$ iterations, and $H$ is the observed image on the ACIS detector.
$P_{jk}$ is the probability that a photon emitted in sky $W$ bin $j$ is measured in data space $H$ bin $k$, or $P(H_k|W_j)$.

\subsection{RL with a Spatially Variant PSF}
Previous Chandra image deconvolution approaches \citep[e.g.,][]{thimmappa2020chandra,sobolenko2022ngc} used a simplified approximation for the $P_{jk}$ values, i.e., they used the same PSF for each $j$ bin. 
Here we assume that the PSF changes as a function of the off-axis angle and the roll angle. 
As a consequence, the Chandra RL~algorithm is extended.
The formula for RL$_{\rm{sv}}$ is obtained by rewriting Equation~(\ref{rl_eq}) as 
\begin{equation}
W_{i}^{(r+1)} = W_{i}^{(r)} \sum_k \frac{P_{iik}H_k}{\sum_j P_{jjk}W_{j}^{(r)}}. 
 \label{RL_sv_eq}
\end{equation}
$P_{jjk}$ refers to a PSF at a position of $j$ (first index) 
which returns a probability that an event emitted at $W_j$ (second index) is observed at $H_k$ (third index), or $P_j(H_k|W_j)$. 
Computational cost and memory requirements need to be minimized for calculating the third-order tensor of the PSF, which is a distinctive feature of the RL$_{\rm{sv}}$~algorithm. 
When $H$ is corrected for slight differences among pixels in effective area and exposures, its normalization can be chosen arbitrarily. 
Here we use $H_k=N_k/A_k$, where $N_k$ is the detector count image, and $A_k$ is the Hadamard product of effective area and exposure time.

\subsection{RL\textsubscript{sv} Deconvolution with Total Variation Regularization}\label{RL_sv deconvolution with the TV regularization}
There are regularization techniques to enhance the RL~method, which are also readily available for the RL$_{\rm{sv}}$~method. 
Among these, total variation (TV) regularization \citep{rudin1992nonlinear} is effective in handling statistical errors, 
which is used in the RL~method \citep{dey2006richardson}. 
The formula for RL$_{\rm{sv}}$ with the regularization is obtained by rewriting Equation~(\ref{RL_sv_eq}) as 
\begin{equation}
 W_{i}^{(r+1)} = \frac{W_{i}^{(r)}}{1-\lambda_{\rm{TV}}\textrm{div}\left( \frac{\nabla W_{i}^{(r)}}{|\nabla W_{i}^{(r)}|}\right)}\sum_k \frac{P_{iik}H_k}{\sum_j P_{jjk}W_{j}^{(r)}}. 
 \label{tv_reg_RL_sv}
\end{equation}
The only difference from the RL$_{\rm{sv}}$~algorithm of Equation~(\ref{RL_sv_eq}) is the regularization term of $1-\lambda_{\rm{TV}}\textrm{div}(\nabla W_{i}^{(r)}/|\nabla W_{i}^{(r)}|)$, where $\lambda_{\rm{TV}}$ is the regularization parameter, $\textrm{div}(\cdot)$ is the divergence, and $\nabla W_{i}^{(r)}$ is the gradient of $W_{i}^{(r)}$.
In this paper, we utilize the parameter of $\lambda_{\rm{TV}}=0.002$, as proposed by \cite{dey2006richardson}.

\subsection{Comparison to other methods}
Deconvolution methods require an understanding of their applicability to a practical condition, 
as well as optimization of computation cost and accuracy (for features of various methods see \cite{naik2013comprehensive}). 
The RL method is well studied and has been used to incorporate regularization \citep[e.g.,][]{van2000background,dey2006richardson,yuan2008progressive,yongpan2010improved} and the recent trend of deep learning \citep[e.g.,][]{9190825}.
For the Chandra users, RL~deconvolution with a single PSF is frequently used because the method is already implemented as \texttt{arestore} in the Chandra Interactive Analysis of Observations \citep[CIAO;][]{fruscione2006ciao}, Chandra's standard data processing package. 

\begin{figure}[ht!]
 \centering
 \includegraphics[width=1\linewidth]{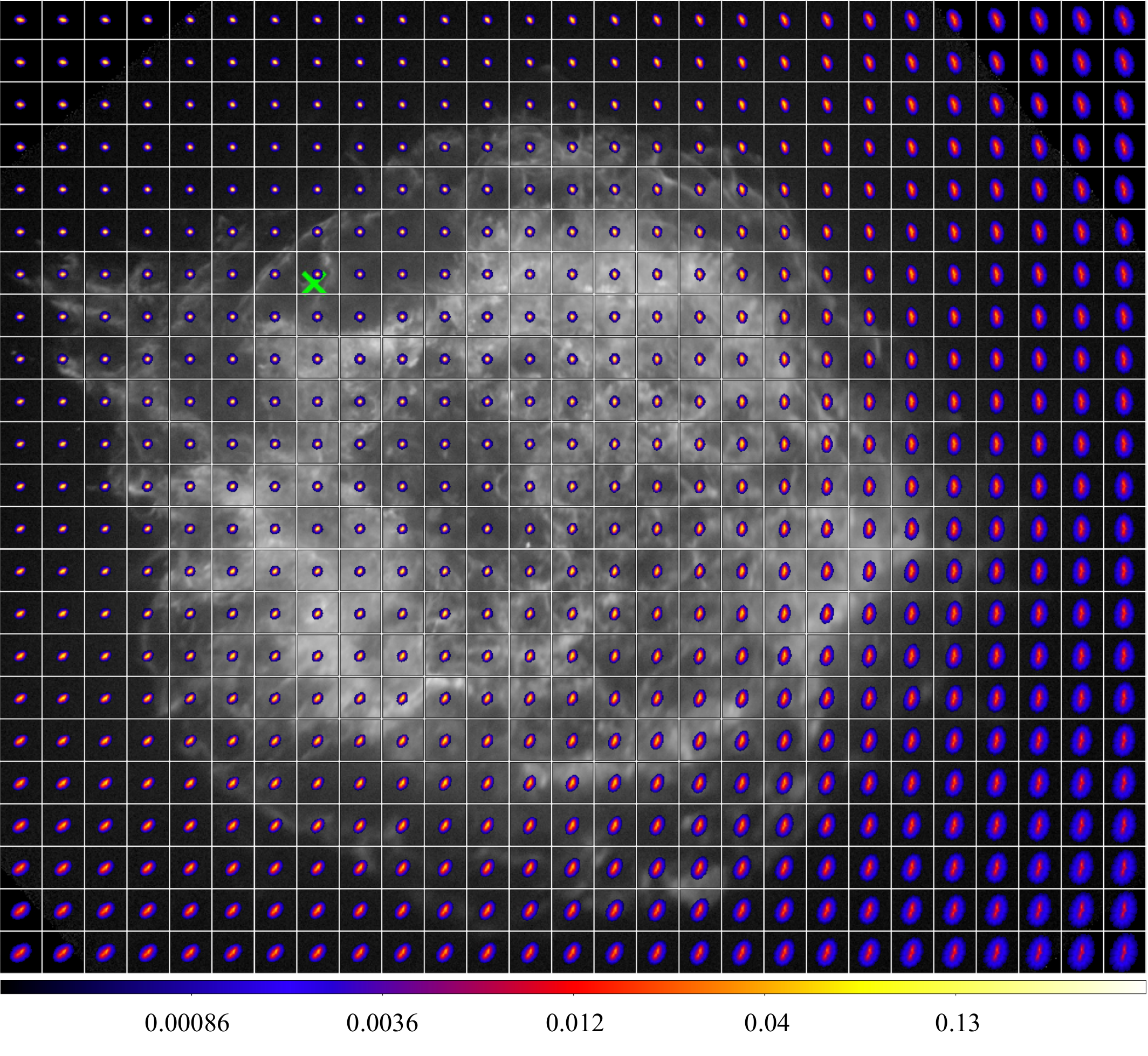} 
 \caption{Cas~A image (Obs.~ID=4636) and the two-dimensional probabilities of the point-spread functions (PSFs). The integral of each PSF is normalized to be 1. The PSF color scale is a fixed range. The location of the optical axis is indicated with a green cross. }
 \label{psf_35bin}
\end{figure}

Compared to other methods, the RL~method forces the deconvolved image of each iteration to be non-negative, and its integral value is conserved. 
Additionally, the method converges to the maximum likelihood solution for a Poisson noise distribution \citep{shepp1982maximum}, which is suitable for Chandra images with noise from counting statistics. 
Depending on the application, it is less prone to ringing artifacts than inverse PSF-based methods \citep[e.g.,][]{sekko1999deconvolution,neelamani2004forward}; see the results of the comparison by \cite{dalitz2015point}. 
According to \cite{white1994image}, it is robust against small errors in the PSF.

\begin{figure*}[ht!]
 \centering
 \includegraphics[width=1\linewidth]{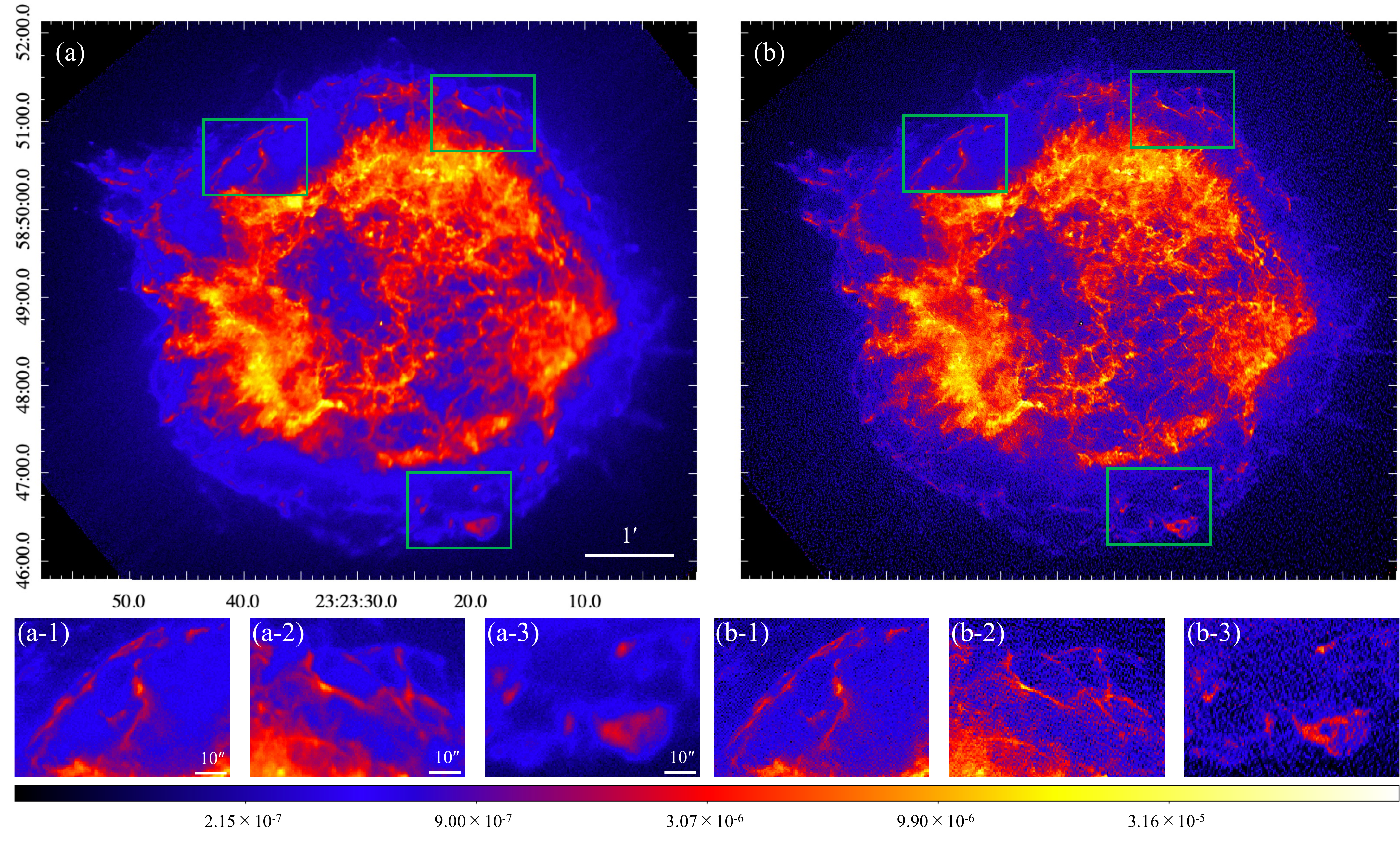}
 \caption{(a) X-ray image in the 0.5--7.0 keV band of Cas~A obtained with Chandra. (a-1, 2, 3) Enlarged images specified by the colored frames in (a). (b) Same as (a), but for the RL$_{\rm{sv}}$-deconvolved results. The unit of flux in the images is $\rm{photons~cm^{-2}~s^{-1}}$.}
 \label{casA_original_rl}
\end{figure*}

\section{Application to Observed Data}\label{Application to Observed Data}
\subsection{Data selection}\label{Data selection}
 
Because Cas~A is a bright and diffuse X-ray source with a moderately large apparent diameter,  it is an ideal target to demonstrate the RL$_{\rm{sv}}$ method. 
It has been observed by Chandra almost every year since 1999. 
The Chandra data of Cas~A used in this paper are 
listed in Table \ref{casA_data}: ACIS-S observation of 2004 in Obs.~ID=4636, 4637, 4639, and 5319.   
The image size is $1489\times 1488$~pixels, or $743''\times 742''$ given a unit pixel of $0.''492$. 
Data processing and analysis were performed using CIAO version 4.13. 
The data were reprocessed from the level 1 event files by \texttt{chandra\_repro}. 
Since the roll angle and optical axis of the four observations are almost the same (the maximum difference of the optical axis location is about 4 unit pixels), 
all the events were merged into one by \texttt{merge\_obs}.
The total exposure time was 428.29 ks.

\subsection{Generating the PSF of Chandra}
The Chandra telescope system consists of four pairs of nested reflecting surfaces, configured in the Wolter type I geometry. 
The high energy response is achieved by coating the mirrors with iridium. 
It has attained the highest angular resolution of $0.''492$ among existing X-ray telescopes.
Its mirror of Chandra has been extensively calibrated on the ground and in orbit \citep{jerius2000orbital}. 
The Chandra PSF is positional-dependent, mainly due to aberrations.
The RL$_{\rm{sv}}$~method includes the position dependence of the PSF, which is useful for highly extended X-ray sources. 

Because creating a PSF for each position is computationally expensive, it is decimated at some intervals. 
For reference, creating a PSF takes several seconds, depending on the computational environment and desired accuracy.
The sampling interval of PSFs was chosen to be $35\times 35$~pixels (total of $43\times 43=1849$). 
The interval was determined empirically by trying several different ranges.
In general, the PSFs simulated from each observation should be merged for a precise calculation. 
Here, the PSF of the Obs.~ID of 4636 was used as a representative since its sampling of the PSF is decimated.  
The PSFs at the lattice points were generated by CIAO's \texttt{simulate\_psf} using the Model of AXAF Response to X-rays \citep{wise1997simulated,davis2012raytracing} at a monochromatic energy of 2.3 keV. 
They were applied to the observed image with energies from 0.5 to 7.0 keV. 

Figure~\ref{psf_35bin} shows all the PSFs sampled every $35\times 35$~pixels. 
The optical axis is located at the northeast in the image, where the spread of the PSF is minimum. 
As the position is away from the optical axis, the tail of the PSF increases with its gradual shift of the elliptical axis. 
Although it is a trade-off with photon statistics, it is effective to run the RL$_{\rm{sv}}$~method with the optimal monoenergetic PSF for each of the multiple energy decompositions (see Figure~\ref{casA_original_rl_rgb}). 
This is because, at shorter wavelengths, the effect of diffuse reflection due to the roughness of the mirror surface is not negligible \citep{jerius2000orbital}.

\subsection{Results of the RL\textsubscript{sv}~method}\label{Results of RL_sv method}
Figure~\ref{casA_original_rl}(a) is an observed $\sim$400~ks image using the energy range 
from 0.5 to 7.0 keV, as explained in Section~\ref{Data selection}. 
We applied the RL$_{\rm{sv}}$~method to the image with a sampling interval of each PSF of $35\times 35$~pixels.
The number of iterations is 200. 
Note that the choice of the iteration number is discussed in Section~\ref{Assessment of the reasonable number of iterations}.
The result of the RL$_{\rm{sv}}$~method is presented in Figure~\ref{casA_original_rl}(b).
The unit of flux and its range in Figure~\ref{casA_original_rl}(b) is the same as in Figure~\ref{casA_original_rl}(a). 
The overall structures in the RL$_{\rm{sv}}$~image become more vivid than the original ones. 
The images around the off-axis are significantly improved compared to those around the optical axis.

To make the differences more precise, 
we present magnified images of the original image in Figures~\ref{casA_original_rl}(a-1,~2,~3) and the RL$_{\rm{sv}}$-image in Figures~\ref{casA_original_rl}(b-1,~2,~3). 
The three regions represent a sharp filament in the northeast, complicated filaments in the north, and a slightly diffuse area in the south. 
The filamentary structures in the northeast and north become sharper in the RL$_{\rm{sv}}$-image. 
We will quantify the filament width in detail and discuss the systematic uncertainties associated with the method in Section~\ref{Assessment of image blurredness}. 

\section{Uncertainty Estimation}\label{Uncertainty Estimation}
\subsection{Assessment of the reasonable number of iterations}\label{Assessment of the reasonable number of iterations}
We considered a way of assessing an appropriate number of iterations, which is one of the issues with the RL~method. 
This is because the method has the property of excessive amplification of noise as the number of iterations increases. 
We propose a method to suppress convergence using statistical errors during the iteration. 
The formula of the method is written as 
\begin{equation}
 W_{i}^{(r+1)} = W_{i}^{(r)}\sum_k \frac{P_{iik}G(N_k)/A_k}{\sum_j P_{jjk}W_{j}^{(r)}}. 
 \label{statical_RL_sv_eq}
\end{equation}
The only difference from the RL$_{\rm{sv}}$~algorithm, Equation~(\ref{RL_sv_eq}), is the $G(N_k)/A_k$ term. 
$N$ ($\rm{counts}$) is the map of detector counts.
$A$ ($\rm{photons~cm^{-2}~s^{-1}}$) is the Hadamard product of effective area and exposure time. 
$G(N_k)$ is a random number generator following a Poisson distribution with a count in the $k$th pixel of $N_k$; i.e., 
$G(N_k)/A_k$ is a flux in units of $\rm{photons~cm^{-2}~s^{-1}}$.
The reason for normalizing $G(N)$ by dividing it with $A$ is to account for the slight variations in effective area and exposure time among pixels.

\begin{figure}[ht!]
 \centering
 \includegraphics[width=1\linewidth]{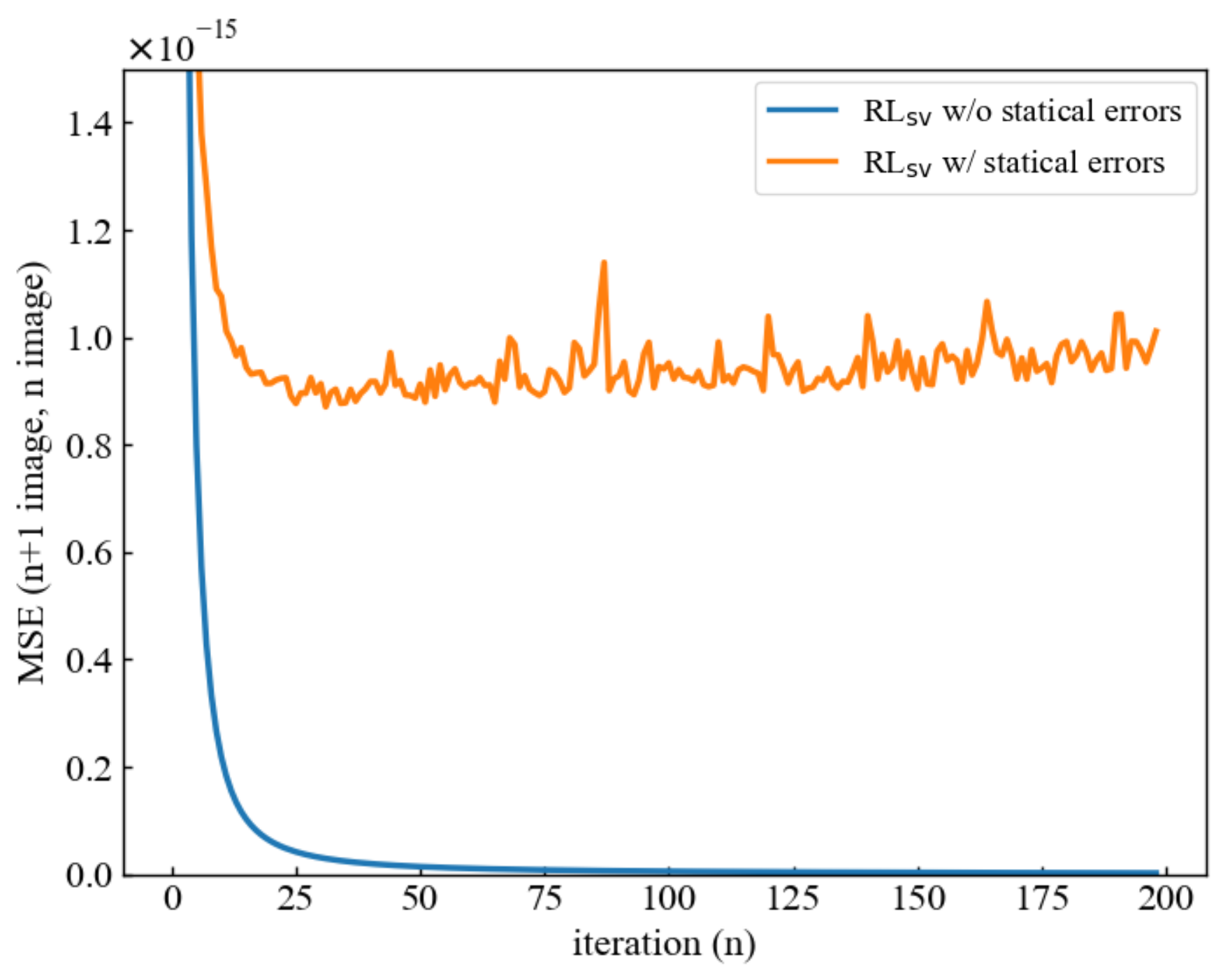}
 \caption{Residuals of the two images before and after iterations for the entire region of Cas~A vs. the number of iterations. The results of RL$_{\rm{sv}}$~method with and without statistical errors are plotted as a blue and an orange line, respectively.}
 \label{normal_and_poisson_err}
\end{figure}

The performance of the RL$_{\rm{sv}}$~algorithm using Equation~(\ref{statical_RL_sv_eq}) is compared to that using Equation~(\ref{RL_sv_eq}). 
The convergence is evaluated by using the mean squared error (MSE) of the two images: one step before and after iterations. 
Figure~\ref{normal_and_poisson_err} shows the history of the MSE during the iteration.  
The curve obtained by the RL$_{\rm{sv}}$~algorithm, Equation~(\ref{statical_RL_sv_eq}) saturates at a certain level, 
while the other continues to decrease.  
The saturation level is caused by the injection of Poisson fluctuation at each step, 
which is considered as an indicator of stopping. 
In Figure~\ref{normal_and_poisson_err}, the iteration number of $\sim$30 seems appropriate. 

\subsection{Assessment of image blurredness}\label{Assessment of image blurredness}
We then designed a simplified method for evaluating a certain amount of confidence. 
The RL$_{\rm{sv}}$-deconvolved image should have a similar amount of fluctuation accompanying the observation image.
The principle of the method is to propagate errors of the converged RL$_{\rm{sv}}$~image into the next step. 
Our choice of using the last step for the error propagation is just to simplify the task. 
Here, each error in the observed image is considered statistically independent.
Assuming only uncertainties on $H_k$, using the law of error propagation, the image uncertainty can be expressed as
\begin{equation}
\begin{split}
 \sigma_{W^\prime_i} &= \sqrt{\sum_k\left[\frac{\partial}{\partial H_k}\left(W_{i}\sum_k \frac{P_{iik}H_k}{\sum_j P_{jjk}W_{j}}\right)\sigma_{H_k}\right]^2}\\
 &= W_{i}\sqrt{\sum_k \left(\frac{P_{iik}}{\sum_j P_{jjk}W_{j}}\frac{\sqrt{N_k}}{A_k}\right)^2}, 
\end{split}
\label{error propagation}
\end{equation}
where $W^\prime$ is the image of the next iteration number of any estimated true image of $W$, and $\sqrt{N_k}$ is the statistical error of $N_k$.

\begin{figure}[ht!]
 \centering
 \includegraphics[width=1\linewidth]{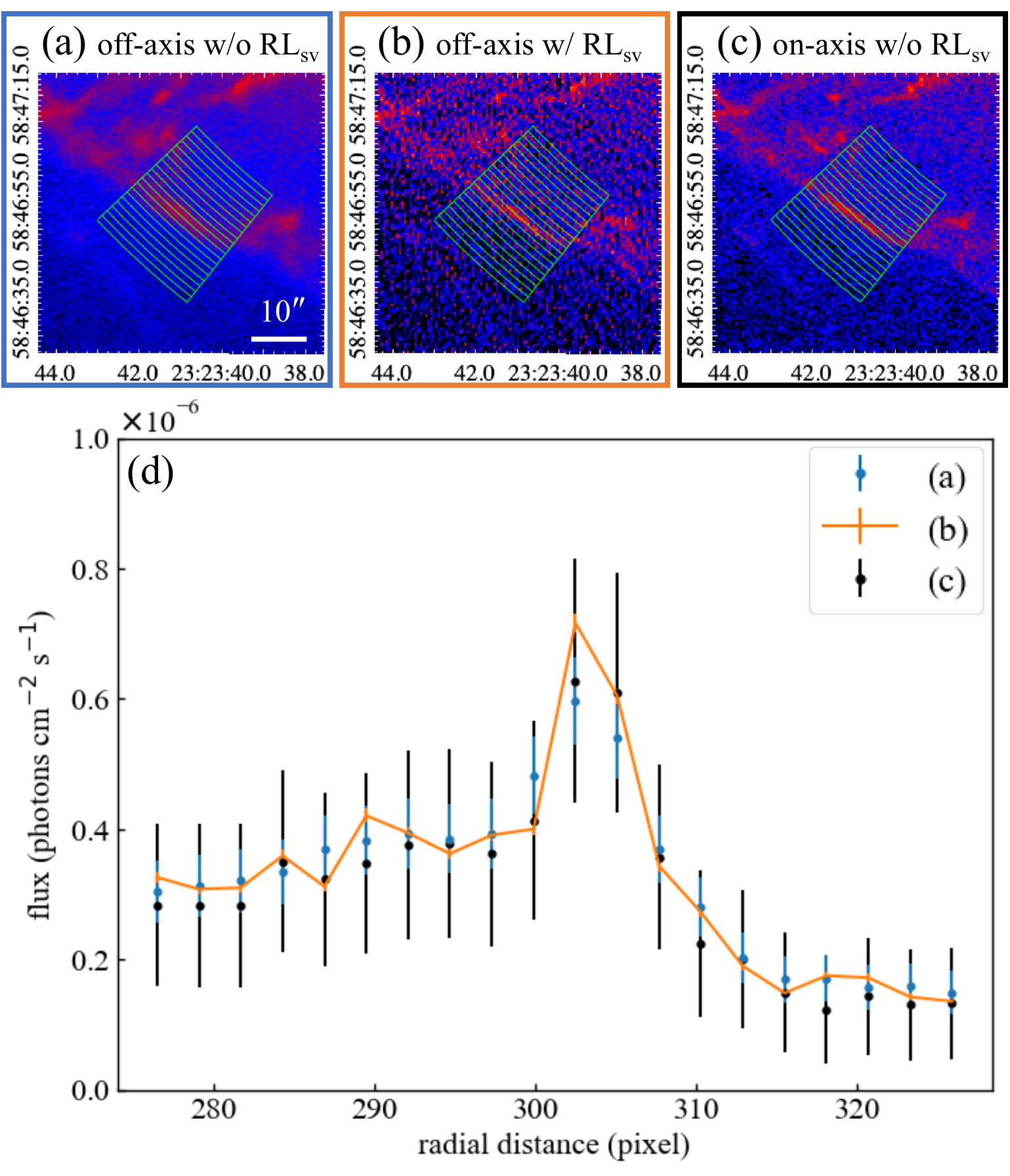}
 \caption{Images and radial profiles of the southeastern filament of Cas~A. (a) Off-axis image of Obs.~ID=4636, 4637, 4639, and 5319. (b) Result of RL$_{\rm{sv}}$-image of (a). (c) On-axis image of Obs.~ID=5196. (d) Results of the radial profile in (a), (b), and (c), radially projected from he central compact object (CCO) using the fan-shaped regions in green. The horizontal axis represents the distance from the CCO.}
 \label{rprofile_on_vs_off_axis}
\end{figure}

We compared the off-axis RL~image with the error of Equation~(\ref{error propagation}) to an on-axis observation in Figure~\ref{rprofile_on_vs_off_axis}.
Figures~\ref{rprofile_on_vs_off_axis}(a) and (b) are the off-axis southeastern images from Figures~\ref{casA_original_rl}(a) and (b), respectively. 
Figure~\ref{rprofile_on_vs_off_axis}(c) shows a southeastern on-axis image of Obs.~ID=5196.
The exposure time for the on-axis observation is $\sim$50 ks, resulting in a larger statistical error compared to the off-axis observation of $\sim$400 ks.
The fan-shaped regions in Figures~\ref{rprofile_on_vs_off_axis}(a)--(c), along the filament, are chosen to create the radial profiles, 
which is the one-dimensional profile of the photons in each region extending from the central compact object (CCO) of Cas~A toward the outer regions. 
These radial profiles were created using \texttt{dmextract} in CIAO.
In Figure~\ref{rprofile_on_vs_off_axis}(d), the framed regions from Figures~\ref{rprofile_on_vs_off_axis}(a)--(c) are color-coded as blue, orange, and black, respectively. 
The error bars of the radial profiles in Figure \ref{rprofile_on_vs_off_axis}(d) correspond to statistical errors, represented by blue and black, and the result obtained by applying Equation~(\ref{error propagation}) to the 199th iteration of the RL$_{\rm{sv}}$~image, is indicated by orange.
From Figure~\ref{rprofile_on_vs_off_axis}(d), the profile of the off-axis RL$_{\rm{sv}}$ image agreed with that of the on-axis image within the statistical errors. 
This method gives a guideline for a certain level of confidence associated with the RL$_{\rm{sv}}$~method. 

\section{Discussion}
\subsection{Enhancement Technique and Possibilities}\label{Enhancement Technique and Possibilities}
In this section, further enhancements to the RL$_{\rm{sv}}$~method are discussed. 
The first is to reduce the loss of down-sampling PSFs. 
The positional dependence of the PSF does not contain high-frequency components, 
so the decimation of PSF sampling should work to some extent. 
For a small image such as a core plus jet structure in an active galactic nucleus, keeping a high sampling rate of the PSFs might be possible. 
However, for a largely extended source such as a supernova remnant or galaxy cluster, 
to minimize the sampling rate is critically important for practical use. 
The higher the decimation, the more emphasized the boundary of the segment. 
Taking Cas~A as an example, the edges of specific segments clearly appear when the sampling interval is $35\times 35$~pixels. 
To smooth out the edges, we propose that the PSFs' boundaries be randomly selected from nearby PSFs (see more details in the \hyperref[Boundaries of the PSF]{Appendix}). 
\begin{figure}[]
 \centering
 \includegraphics[width=1\linewidth]{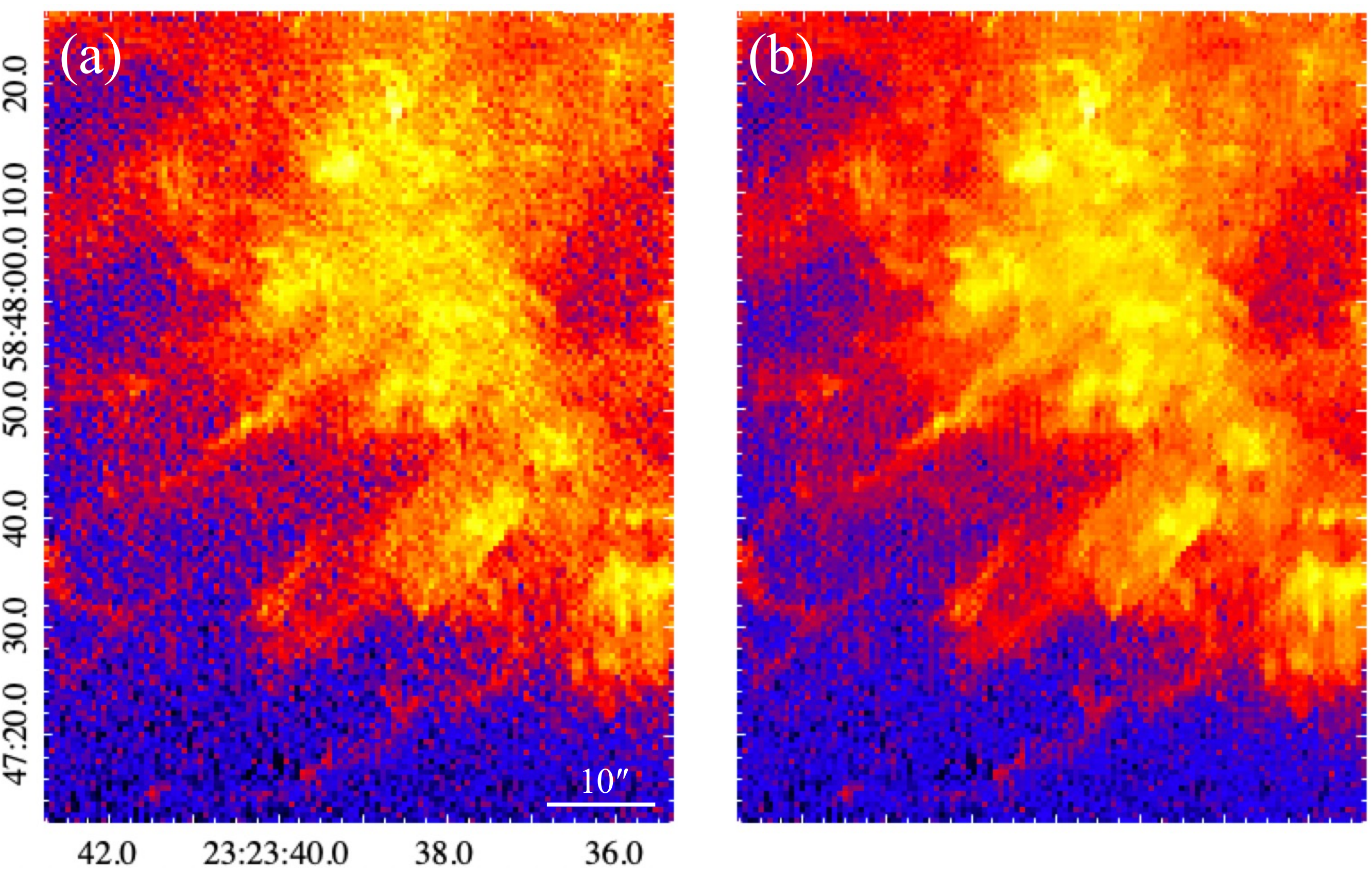}
 \caption{Comparison of the results of the RL$_{\rm{sv}}$~method without and with total variation regularization, shown in (a) and (b) respectively.}
 \label{tv_image}
\end{figure}

Second, this method can be developed by incorporating several regularization methods.
We implemented an RL$_{\rm{sv}}$~method incorporating the TV regularization expressed in Equation~(\ref{tv_reg_RL_sv}).
Finally, the RL$_{\rm{sv}}$ method is naturally applied to color images. 
By decomposing observed images into several colors (or energy bands) and generating PSFs for an appropriate energy in each band, 
an energy-dependent RL$_{\rm{sv}}$~method can be realized.

\begin{figure*}[ht!]
 \centering
 \includegraphics[width=1\linewidth]{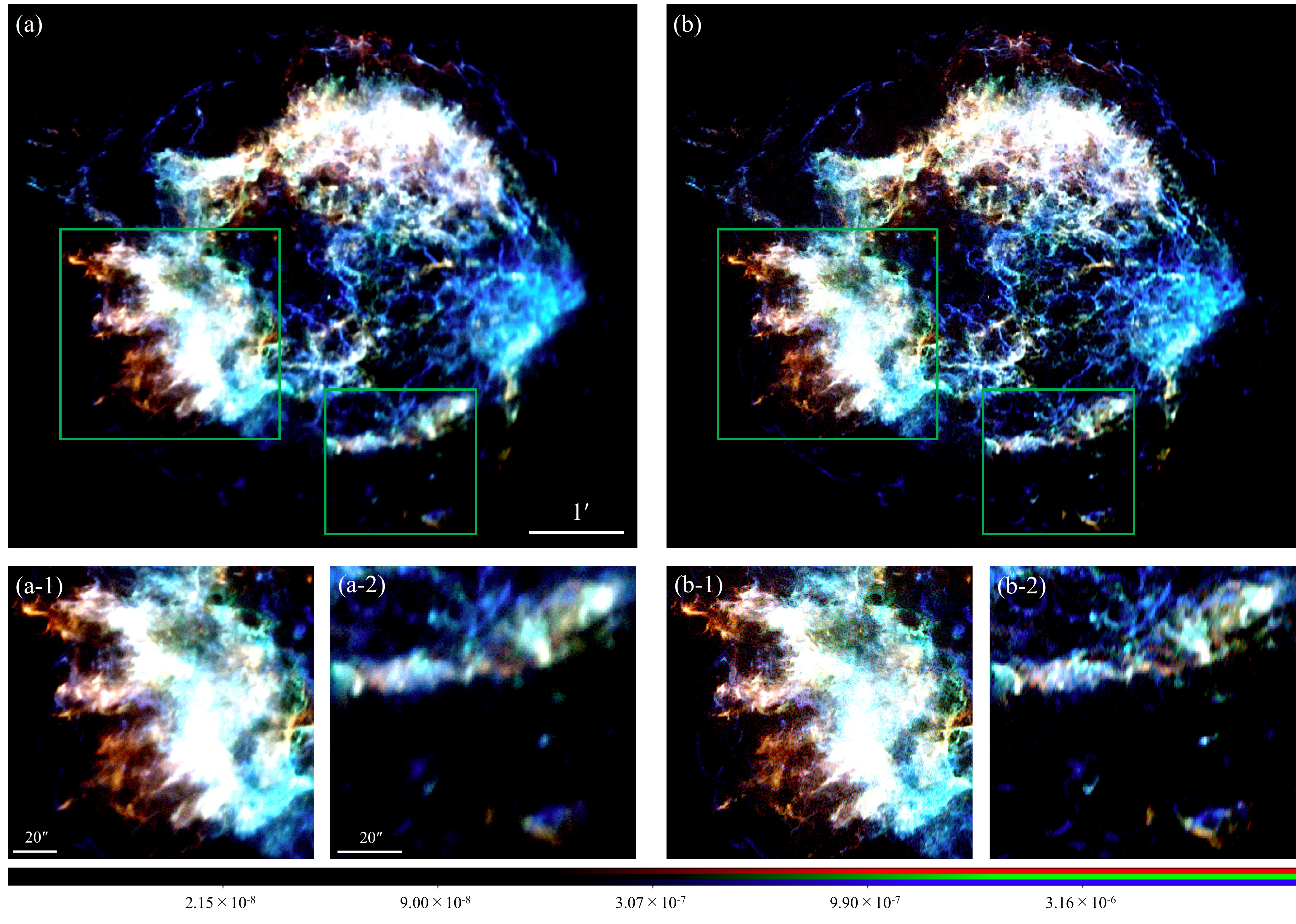}
 \caption{(a) X-ray RGB (red: 0.2--1.2 keV, green: 1.2--2.0 keV and blue: 2.0--7.0 keV) band images of Cas~A obtained with Chandra. (a-1, -2): Enlarged images specified by the colored frames in (a). (b) Same as (a) except for RL$_{\rm{sv}}$-deconvolved in each energy band. The unit of flux in the images is $\rm{photons~cm^{-2}~s^{-1}}$.}
 \label{casA_original_rl_rgb}
\end{figure*}

We compare the RL$_{\rm{sv}}$~method with and without the TV regularization.
We use the same image as in Section~\ref{Data selection}.
The PSF sampling is $35\times 35$ pixels and the number of iterations is 200. 
The PSFs' boundaries are randomly selected following the \hyperref[Boundaries of the PSF]{Appendix}.
Figure~\ref{tv_image} presents the enlarged eastern image after applying the RL$_{\rm{sv}}$ method to the entire region. 
Figures~\ref{tv_image}(a) and (b) show the results of the RL$_{\rm{sv}}$~method of Equation~(\ref{RL_sv_eq}) and the regularization version of Equation~(\ref{tv_reg_RL_sv}), respectively.
The TV regularization preserves the sharp structure to remain and smoothes out statistical errors.
In this way, regularization can be added to the RL$_{\rm{sv}}$~method.

We implement the RL$_{\rm{sv}}$~method including these enhancements and adapt it to the Cas~A observational data described in Section~\ref{Data selection}. 
Figure~\ref{casA_original_rl_rgb}(a) is the observed image of Section~\ref{Results of RL_sv method} divided into three energies in RGB: 0.2--1.2 keV (red), 1.2--2.0 keV (green), and 2.0--7.0 keV (blue). 
Cas~A is dominated by the thermal radiation in $\leqslant$4 keV and the nonthermal radiation in $\geqslant$4 keV.
We applied the RL$_{\rm{sv}}$~method with the TV regularization Equation~(\ref{tv_reg_RL_sv}) to each energy image of Figure~\ref{casA_original_rl_rgb}(a) using the appropriate energy of the PSF (red is 0.92 keV, green is 1.56 keV, and blue is 3.8 keV) based on the official CIAO page.\footnote{\url{https://cxc.cfa.harvard.edu/ciao/why/monochromatic_energy.html}}
The sampling interval of PSFs is $35\times 35$ pixels.
PSF is randomly selected at the sampling boundaries according to the \hyperref[Boundaries of the PSF]{Appendix}. 
The number of iterations is 30, according to Section \ref{Assessment of image blurredness}. 
The result of the RL$_{\rm{sv}}$~method is presented in Figure~\ref{casA_original_rl_rgb}(b). 
The energy dependence in Figures~\ref{casA_original_rl_rgb}(b-1) and (b-2) are clearly visible by this method.

\subsection{Constraint on the Uncertainty}
We propose two complementary ways to obtain a guideline on the stop condition of the RL$_{\rm{sv}}$~method.
One is to obtain a minimum of residuals by inserting statistical uncertainties into each update during the iteration process.
This gives a rough estimate of the limit of the iterations.
The other is to include the statistical uncertainties in the last step of the iteration.
By combining the two methods, it is possible to derive uncertainties using the errors obtained by the latter method at the optimal iteration number estimated by the former.
This is a quick and convenient way to derive the systematic uncertainties associated with the RL$_{\rm{sv}}$~method.

The systematic uncertainty in the former method is a way of defining the optimal number of iterations. 
One easy way is to use the same level of residuals as shown in Section~\ref{Assessment of the reasonable number of iterations}.
It is intrinsically difficult to distinguish the signal of the celestial objects from the statistical noise.
This difficulty needs to be overcome by comparing the deconvolved images without errors and the error-estimated images around the optimal iteration number recommended by the former method. 
This method is based on a compromise between the computational cost and the simplicity of use while keeping a reasonable statistical error. 

\section{Conclusion}
We have improved the processing capability of RL deconvolution by incorporating the positional dependency of the Chandra PSF. 
The RL$_{\rm{sv}}$ method is applied to the entire region of Cas~A with an estimation of its limit and errors, 
which are based on the phenomenological method for evaluating a reasonable number of iterations and uncertainties. 
It shows that the features of shock waves and jets are sharper than those measured in the original image, with a certain amount of knowledge of the associated errors. 
The RL$_{\rm{sv}}$-deconvolved profile of the off-axis image at the southeastern filament became shaper and agreed with that of the on-axis observation within the statistical errors.
This method is useful for a detailed diagnosis of other extended X-ray sources obtained by Chandra. 
\newline \par
The code used in this paper is available at doi:10.5281/zenodo.8020557.
\newline \par
We would like to thank the anonymous referee for helpful comments and feedback on this paper. This research has made use of data obtained from the Chandra Data Archive and the Chandra Source Catalog, and software provided by the Chandra X-ray Center (CXC) in the application packages CIAO. 
This work was supported by JSPS KAKENHI grant Nos. 20K20527, 22H01272, and 20H01941. 

\section*{Appendix\\ Boundaries of the PSF} \phantomsection \label{Boundaries of the PSF}
\begin{figure}[ht!]
 \centering
 \includegraphics[width=1\linewidth]{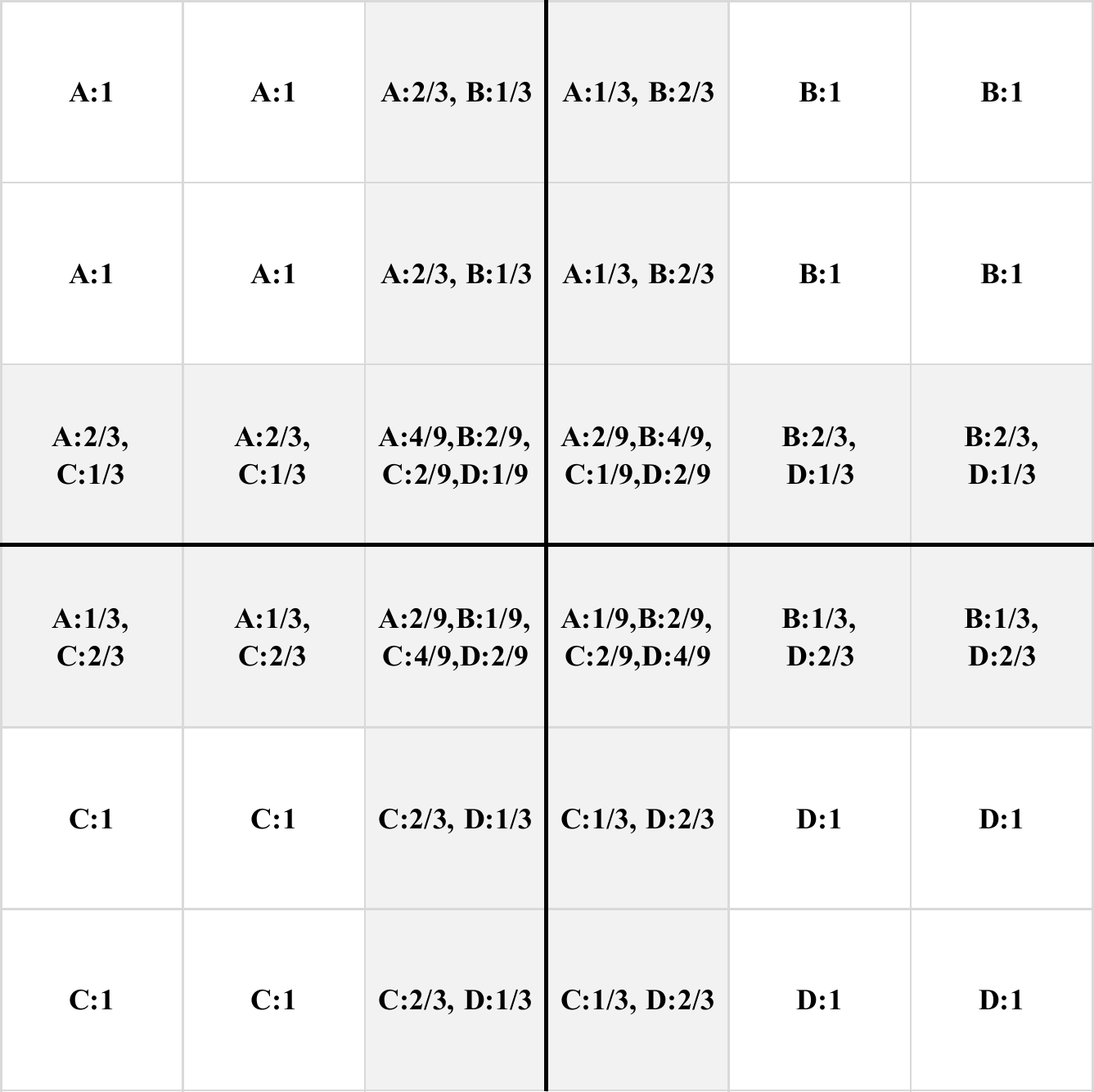}
 \caption{Illustration of the $9\times 9$~pixels around the intersection of the PSF switchover, overlaid with the probability weights on selecting PSFs. 
The quadrants are named A, B, C, and D for illustrative purposes. 
The weights of the probabilities are either 1/3 or 2/3 at the two boundaries, while they are 1/9, 2/9, or 4/9 at the corners.}
 \label{psf_grid_change}
\end{figure}

Decimating the sampling number of PSFs is an effective approach to minimize computational cost.
However, this technique can introduce side-effects at the boundary of segments when switching between PSFs.
The variation in shape between neighboring PSFs, caused by the sampling interval, leads to discontinuities in the deconvolution process. 
To mitigate this issue, a simple countermeasure is to randomly select adjacent PSFs at their boundaries, which helps to smooth out the discontinuities. 
The weights of the probabilities for selecting PSFs are illustrated in Figure~\ref{psf_grid_change}.
The presence and severity of artifacts depend on factors such as the dissimilarity in shape between neighboring PSFs, statistical characteristics of the observed images, and other relevant factors. 
Therefore, the presented technique serves as an example, and the problem is optimized by the range of pixels to be randomized.

\begin{figure}[ht!]
 \centering
 \includegraphics[width=1\linewidth]{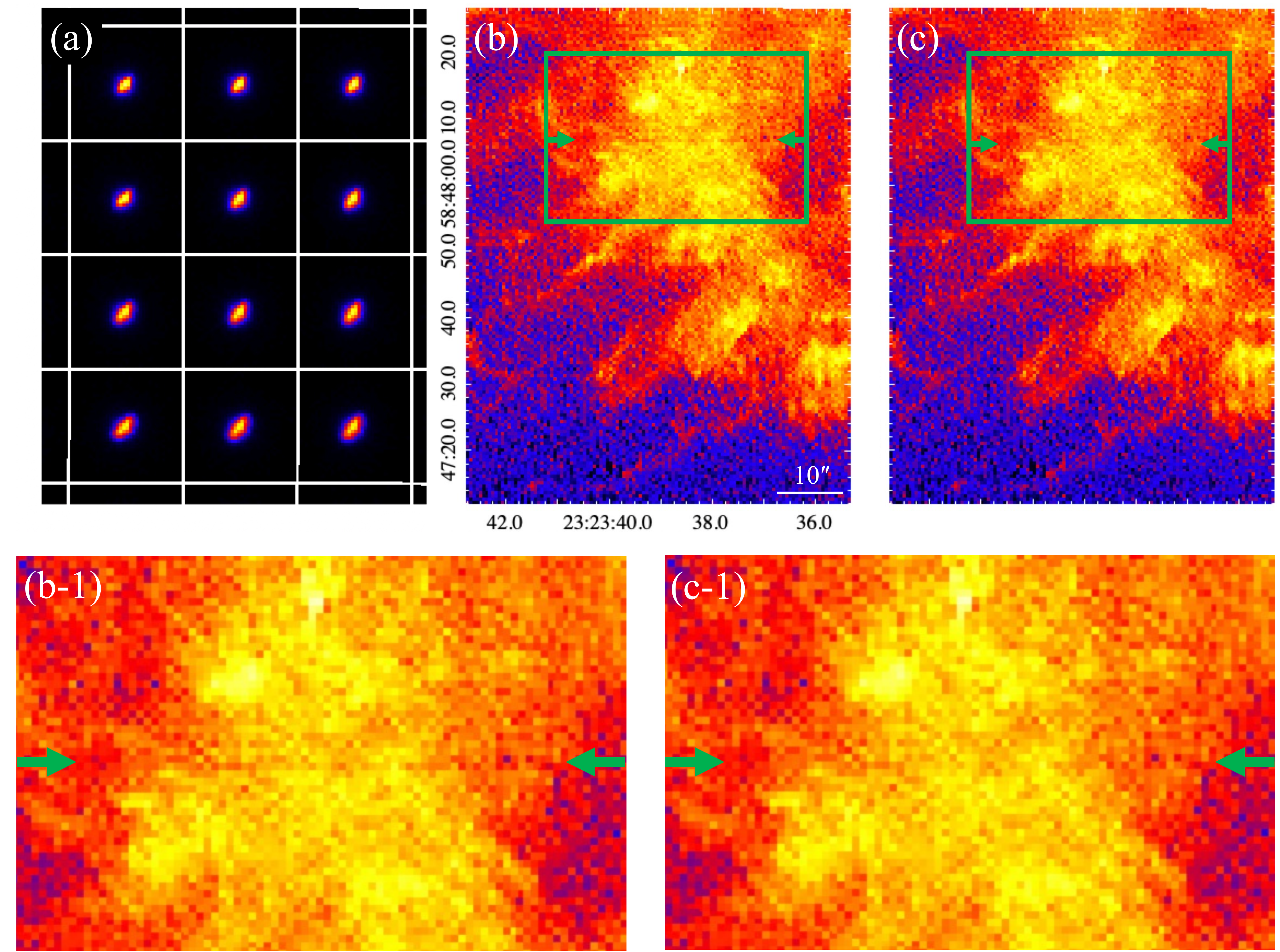}
 \caption{Comparison of the RL$_{\rm{sv}}$~method without and with PSF randomization at the boundaries. (a) PSF images corresponding to (b) and (c). (b) RL$_{\rm{sv}}$-image without correcting the PSF boundaries. (b-1) Enlarged images specified by the colored frames in (b). The arrows correspond to the boundaries of the PSFs. (c) Same as (b), but using the randomization of the PSFs.}
 \label{edge_smoothing_sample}
\end{figure}

The result of applying the selection rule in Figure~\ref{psf_grid_change} is shown in Figure~\ref{edge_smoothing_sample}.
We compared the RL$_{\rm{sv}}$ method for the observed data in the eastern region in Section~\ref{Data selection} with a PSF sampling of $35\times 35$ pixels and 200 iterations.
Figure~\ref{edge_smoothing_sample}(a) shows the PSF images corresponding to Figures~\ref{edge_smoothing_sample}(b) and (c), 
where the white lines are used to clarify the border lines.
Figures~\ref{edge_smoothing_sample}(b) and (c) are the RL$_{\rm{sv}}$-deconvolved images with and without using the randomization of PSFs, respectively. 
To illustrate the differences, we presented magnified images of Figures~\ref{edge_smoothing_sample}(b) and (c) as Figures~\ref{edge_smoothing_sample}(b-1) and (c-1), respectively. 
It appears that the discontinuity at the boundaries of the PSFs, indicated by the green arrows, is smeared out to some extent.

\bibliography{RL_sv_apj}{}
\bibliographystyle{aasjournal}

\end{document}